\documentclass[universe,article,submit,pdftex]{Definitions/mdpi}
\usepackage{amsmath,latexsym}
\nolinenumbers

\firstpage{1}
\pubvolume{xx}
\issuenum{1}
\articlenumber{0}
\pubyear{2021}
\copyrightyear{2021}
\history{Received: ; Accepted: ; Published: date}

\Title{Effects of Higher Order Retarded Gravity}
\Author{Asher Yahalom $^{1,2}$\orcidA{}}

\AuthorNames{Asher Yahalom}

\address{%
$^{1}$ \quad Ariel University, Faulty of Engineering, Department of Electrical \& Electronic Engineering, Ariel 40700, Israel; asya@ariel.ac.il\\
$^{2}$ \quad Ariel University, Center for Astrophysics, Geophysics, and Space Sciences (AGASS),
 Ariel 40700, Israel;}

\abstract{In a recent paper we have a shown that the flattening of galactic rotation curves can be explained by retardation. However, this will rely on a temporal change of galactic mass. In our previous work we have kept only second order terms of the retardation time in our analysis, while higher terms in the Taylor expansion where not considered. Here we consider analysis to all orders and show that indeed a second order analysis will suffice, and higher order terms can be neglected. }


\begin{document}
\nolinenumbers

\newcommand{\beq} {\begin{equation}}
\newcommand{\enq} {\end{equation}}
\newcommand{\ber} {\begin {eqnarray}}
\newcommand{\enr} {\end {eqnarray}}
\newcommand{\eq} {equation}
\newcommand{\eqs} {equations }
\newcommand{\mn}  {{\mu \nu}}
\newcommand{\abp}  {{\alpha \beta}}
\newcommand{\ab}  {{\alpha \beta}}
\newcommand{\sn}  {{\sigma \nu}}
\newcommand{\rhm}  {{\rho \mu}}
\newcommand{\sr}  {{\sigma \rho}}
\newcommand{\bh}  {{\bar h}}
\newcommand{\br}  {{\bar r}}
\newcommand {\er}[1] {equation (\ref{#1}) }
\newcommand {\ern}[1] {equation (\ref{#1})}
\newcommand {\Ern}[1] {Equation (\ref{#1})}
\newcommand{\hdz}  {\frac{1}{2} \Delta z}

\section {Introduction}

Einstein's general relativity (GR) is known to be invariant under general coordinate modifications.
This group of general transformations has a Lorentz subgroup, which is valid even in the weak field approximation. This is seen through the field equations containing the d'Alembert (wave) operator, which can be solved using a {\bf retarded potential~solution}.

It is known that GR is verified by many types of obser\-vations.
How\-ever, curr\-ently, Ne\-wton\-–Eins-\\tein gravitational theory is at a crossroads. It has much in its favor observationally, and it has some very disquieting challenges. The~successes that it has achieved in both astrophysical and cosmological scales have to be considered in light of the fact that GR needs to appeal to two unconfirmed ingredients, dark matter and energy, to achieve these successes. Dark matter has not only been with us since the 1920s (when it was initially known as the missing mass problem), but~it has also become severe as more and more of it had to be introduced on larger distance scales as new data have become available. Here we will be particularly concerned in the excess dark matter needed to justify observed gravitational lensing. Moreover, 40-year-underground and accelerator searches and experiments have failed to establish its existence. The~dark matter situation has become even more disturbing in recent years as the Large Hadron Collider was unable to find any super symmetric particle candidates, the community's preferred form of dark matter.

While things may still take turn in favor of the dark matter hypothesis, the~current situation is serious enough to consider the possibility that the popular paradigm might need to be amended in some way if not replaced altogether. The~present paper seeks such a modification. Unlike other ideas such as Milgrom's MOND \cite{Mond}, Bekenstein's TeVeS \cite{Bekenstein},
Mannheim's Conformal Gravity \cite{Mannheim0,Mannheim1,Mannheim2},
Moffat's MOG \cite{MOG} or $f(R)$ theories and scalar-tensor gravity \cite{Corda}, the~present approach is, the~minimalist one adhering to the razor of Occam. It suggests to replace dark matter by effects within standard GR.

Dynamics of large scale structures is inconsistent with Newtonian mechanics. This was notified in the 1930's by Fritz Zwicky \cite{zwicky},
who pointed out that if more (unseen) mass would be present one would be able to solve the apparent contradiction. The phenomena was also observed in galaxies by Volders \cite{volders} who have shown that star trajectories near the rim of galaxies do not move according to Newtonian predictions, and later corroborated by Rubin and Ford~\cite{rubin1,rubin2,Binney} for spiral galaxies.

In as series of papers we have shown that those discrepancies can be shown to result from retarded gravity as dictated by the theory of general relativity \cite{YaRe1,ge,YaRe2,YahalomSym,Wagman,YaRe3}. Indeed in the absence of temporal density changes, retardation does not effect the gravitational force. However, density is not constant for galaxies, in fact there are many processes that change the mass density in galaxies over time. Mass accretion from the intergalactic medium and internal processes such as super novae leading to super winds \cite{Wagman} modify the density. In addition to those local processes there is a cosmological decrease of density due to the cosmic expansion. However, the later process is many orders of magnitude smaller than the former.

In previous analysis \cite{YaRe1,ge,YaRe2,YahalomSym,Wagman,YaRe3} the corrected gravitational force was evaluated assuming a second order approximation in the retardation time $\frac{R}{c}$, neglecting higher order terms without justification. Here we take into account all orders and show that a second order approximation is indeed sufficient.

\section {Linear GR}

Only in cases of  extreme compact objects (black holes and neutron stars) and the very early universe we consider the solution of the full non-linear Einstein Equations~\cite{YaRe1}. In~most cases of astronomical interest (including the galactic case) a linear approximation to those equations around the flat Lorentz metric $\eta_{\mn}$ is used, such that:
 \beq
 g_{\mn} = \eta_{\mn} + h_{\mn}, \quad \eta_{\mn} \equiv \ {\rm diag } \ (1,-1,-1,-1),
 \qquad
 |h_{\mn}|\ll 1
 \label{lg}
 \enq

 One then defines the quantity:
 \beq
 \bar h_\mn \equiv h_\mn -  \frac{1}{2} \eta_\mn h, \quad h = \eta^{\mn} h_{\mn},
 \label{bh}
 \enq
 $\bar h_\mn = h_\mn $ for non diagonal terms. For~diagonal terms:
 \beq
 \bar h = - h \Rightarrow  h_\mn = \bar h_\mn -  \frac{1}{2} \eta_\mn \bar h .
 \label{bh2}
 \enq

 It~was shown (\cite{Narlikar} page 75, exercise 37, see also~\cite{Edd,Weinberg,MTW}) that for a proper gauge the Einstein equations are:
 \beq
\Box \bh_{\mn} \equiv \bh_{\mn, \alpha}{}^{\alpha}=-\frac{16 \pi G}{c^4} T_\mn , \qquad \bh_{\mu \alpha,}{}^{\alpha}=0.
\label{lineq1}
\enq

\Ern{lineq1} can be solved such that~\cite{padma,Carroll}:
 \ber
& & \bh_{\mn}(\vec x, t) = -\frac{4 G}{c^4} \int \frac{T_\mn (\vec x', t-\frac{R}{c})}{R} d^3 x',
\nonumber \\
 t &\equiv& \frac{x^0}{c}, \quad \vec x \equiv x^a \quad a,b \in [1,2,3], \nonumber \\
  \vec R &\equiv& \vec x - \vec x', \quad R= |\vec R |.
\label{bhint}
\enr

In \cite{Yahalom,Yahalomb,Yahalomc,Yahalomd} we explain why the symmetry between space and time is broken, which justifies the use of different notations for space and time.
$\frac{4 G}{c^4} \simeq 3.3 \times 10^{-44}$; is a tiny number, hence, in~the above calculation
one can take $T_\mn$, to zeroth order in $h_\abp$.{\bf We underline that the term $\frac{R}{c}$ is
not the actual delay time between an arbitrarily moving source and arbitrarily moving observer
and can only interpreted in this way if both are static, nevertheless \ern{bhint} is
an exact solution of \ern{lineq1}. The reader who is interested in the actual delay time is invite to read \cite{Scott,Will}}.
We now evaluate the affine connection in the linear approximation:
\beq
\Gamma^\alpha_\mn = \frac{1}{2} \eta^\abp \left(h_{\beta \mu, \nu} + h_{\beta \nu, \mu} - h_{\mn, \beta}\right).
\label{affinel}
\enq

Notice that the affine connection has first order terms in $h_\abp$; hence, to~the first order
$\Gamma^\alpha_\mn u^\mu u^\nu$ appearing in the geodesic equation,  $u^\mu u^\nu$ must be taken to zeroth order. In~which:
\beq
u^0=\frac{1}{\sqrt{1-\frac{v^2}{c^2}}}, u^a = \vec u =\frac{\frac{\vec v}{c}}{\sqrt{1-\frac{v^2}{c^2}}} ,
\vec v \equiv  \frac{d \vec x}{d t}, \quad v= |\vec v|.
\label{uz}
\enq

For  velocities much smaller than the speed of light in vacuum:
\beq
u^0 \simeq 1,  \qquad \vec u \simeq \frac{\vec v}{c} , \qquad u^a \ll u^0   \qquad {\rm for} \quad v \ll c.
\label{uzslo}
\enq

Hence, the current paper does not discuss the post-Newtonian approximation, in~which matter travels at speeds close to the speed of light, but~
we do consider the retardation effects which are due to the finite propagation speed of the gravitational field. We emphasize that
assuming $\frac{v}{c} << 1$ is not the same as stating $\frac{R}{c} << 1$ (with $R$ being the typical size of a galaxy) since:
\beq
\frac{R}{c} = \frac{v}{c} \frac{R}{v}
\label{retarba}
\enq
now since in~galaxies, $\frac{R}{v}$ is  huge  ($\frac{R}{v} \simeq 10^{15}$ seconds); it follows that, $\frac{v}{c}$ can be dismissed but
not $\frac{R}{c}$, for~which $\frac{R}{c} \simeq 10^{12}$ seconds.
Inserting Equations~(\ref{affinel}) and (\ref{uzslo}) in the geodesic equation, we arrive at the approximate equation:
\beq
\frac{d v^a}{dt}\simeq - c^2 \Gamma^a_{00} = - c^2 \left( h^a_{0,0} - \frac{1}{2} h_{00,}{}^a \right)
\label{geol}
\enq

Taking a standard matter $T_\mn$,~assuming $\rho c^2 \gg p$ and, taking into account Equation~(\ref{uzslo}), we~arrive at $T_{00} = \rho c^2 $, while  the remaining tensor components are much smaller. Therefore, $\bar h_{00}$ is larger than other components of $\bar h_\mn$. Notice that it is not possible to deduce from the magnitudes of quantities that a similar difference
exists between the derivatives of those quantities. Gauge conditions in Equation~(\ref{lineq1}) lead to:
\beq
\bar h_{\alpha 0,}{}^0=-\bar h_{\alpha a,}{}^a \qquad \Rightarrow
\bar h_{00,}{}^0=-\bar h_{0 a,}{}^a, \quad \bar h_{b0,}{}^0=-\bar h_{b a,}{}^a.
\label{gaugeim}
\enq

Thus, the~zeroth derivative of $\bar h_{00}$ (which contains a $\frac{1}{c}$) is of similar order as the spatial derivative
of $\bar h_{0a}$. Also the zeroth derivative of $\bar h_{0a}$ (see equation~(\ref{geol})) is of similar order
as the spatial derivative of $\bar h_{ab}$. However, we can compare spatial derivatives of $\bar h_{00}$ and $\bar h_{ab}$
and conclude that the former is larger. Taking into account equation~(\ref{bh2}) and the above consideration,
we write equation~(\ref{geol}) as:
\beq
\frac{d v^a}{dt}\simeq \frac{c^2}{4} \bar h_{00,}{}^a \Rightarrow \frac{d \vec v}{dt} = - \vec \nabla \phi = \vec F,
\qquad \phi \equiv \frac{c^2}{4} \bar h_{00}
\label{geol2}
\enq

Thus, the gravitational potential $\phi$  can be estimated using Equation~(\ref{bhint}):
\ber
\phi &=& \frac{c^2}{4} \bar h_{00}
= -\frac{ G}{c^2} \int \frac{T_{00} (\vec x', t-\frac{R}{c})}{R} d^3 x'
\nonumber \\
&=& -G \int \frac{\rho (\vec x', t-\frac{R}{c})}{R} d^3 x'
\label{phi}
\enr
and $\vec F$ is the force per unit mass. In the case that the mass density $\rho$ does not depend on time, we may use the Newtonian instantaneous action at a distance. Notice, however, that it is improbable that $\rho$ is static for a galaxy, as~it accretes intergalactic~medium gas. {\bf Also notice that the velocity of galactic and intergalactic matter components (stars, dust \& gas) are implicit in the above formulation as the a time dependent density requires a velocity field
according to the continuity equation (52) of \cite{YahalomSym}.}

Inserting \ern{phi} into \ern{geol2} will lead to:
\ber
\vec F &=& \vec F_{Nr} + \vec F_r
\nonumber \\
 \vec F_{Nr} &=&   -G \int \frac{\rho (\vec x', t-\frac{R}{c})}{R^2} \hat R d^3 x', \qquad \hat R \equiv \frac{\vec R}{R}
\nonumber \\
 \vec F_r &\equiv& - \frac{G}{c} \int  \frac{\rho^{(1)} (\vec x', t-\frac{R}{c})}{R} \hat R d^3 x',
 \qquad \rho^{(n)} \equiv \frac{\partial^n \rho}{\partial t^n}.
\label{Fr}
\enr
Thus a retarded potential does not only imply a retarded Newtonian force $\vec F_{Nr}$, but in addition a pure "retardation" force $\vec F_r$ which decreases slowly than the Newtonian force with distance, explaining the peculiar form of the galactic rotation curves. We emphasize that this result is independent of any perturbation expansion in the delay time $\frac{R}{c}$ as was done in
\cite{YahalomSym}. However, the perturbation expansion does shed some light on the nature of those force terms as will be explained in the next section.

\section {Retardation Effects Beyond the Newtonian Approximation}
\label{REBN}

The duration $\frac{R}{c}$ may be tens of thousands of years, but~may be short with respect
to the duration in which the galactic density changes considerably. Thus, we write a Taylor series for the density:
\beq
\rho (\vec x', t-\frac{R}{c})=\sum_{n=0}^{\infty} \frac{1}{n!} \rho^{(n)} (\vec x', t) (-\frac{R}{c})^n.
\label{rhotay}
\enq

By inserting Equations~(\ref{rhotay}) into Equation~(\ref{phi}), we will obtain:
\ber
\phi &=& \phi_2 + \phi_{(n>2)}
\nonumber \\
\phi_2 &=&  -G \int \frac{\rho (\vec x', t)}{R} d^3 x' +  \frac{G}{c}\int \rho^{(1)} (\vec x', t) d^3 x'
- \frac{G}{2 c^2}\int R \rho^{(2)} (\vec x', t) d^3 x'
\nonumber \\
\phi_{(n>2)} &=& -G \sum_{n=3}^{\infty} \frac{(-1)^n}{n! c^n} \int R^{n-1} \rho^{(n)} (\vec x', t) d^3 x'
\label{phir}
\enr

The Newtonian potential is the first term, the~second term has null contribution, and~the third term is the lower order correction to the Newtonian theory:
\beq
 \phi_r = - \frac{G}{2 c^2} \int  R \rho^{(2)} (\vec x', t) d^3 x'
\label{phir2}
\enq
The~expansion given in Equation~(\ref{phir}), being a Taylor series expansion, is only valid for limited~radii determined by the convergence of the infinite sum:
\beq
R < c \ T_{max} \equiv R_{max}
\label{Rmax}
\enq
hence the current approximation can only be used in the near field regime, this is to be contrasted with the far field approximation used for gravitational radiation \cite{Einstein2,Taylor,Castelvecchi}. The restriction is even more severe when one uses a second order expansion as was done in \cite{YahalomSym}.

If $n>2$ terms can be neglected the total force per unit mass can be approximated as:
\ber
\vec F &\simeq& \vec F_N + \vec F_{ar}
\nonumber \\
 \vec F_N &=& - \vec \nabla \phi_N =  -G \int \frac{\rho (\vec x',t)}{R^2} \hat R d^3 x', \qquad \hat R \equiv \frac{\vec R}{R}
\nonumber \\
 \vec F_{ar} &\equiv& - \vec \nabla \phi_r =  \frac{G}{2 c^2} \int  \rho^{(2)} (\vec x', t) \hat R d^3 x'.
\label{Fr2}
\enr
In the above $\vec F_N$ is a {\bf non retarded} Newtonian force. To see how this comes about from
the existence of a Newtonian retarded force and retardation force as defined in \ern{Fr} we write
those expressions up to order $O(1/c^2)$ using \ern{rhotay}, we thus have:
\ber
 \vec F_{Nr} &\simeq&   -G \int \frac{\rho (\vec x', t)}{R^2} \hat R d^3 x' +
 \frac{G}{c}\int \frac{\rho^{(1)} (\vec x', t)}{R} \hat R d^3 x' -
 \frac{G}{2 c^2} \int \rho^{(2)} (\vec x', t) \hat R d^3 x',
\nonumber \\
 \vec F_r &\simeq& - \frac{G}{c} \int  \frac{\rho^{(1)} (\vec x', t)}{R} \hat R d^3 x' +
 \frac{G}{c^2} \int \rho^{(2)} (\vec x', t) \hat R d^3 x'.
\label{Fr3}
\enr
Adding those two terms we see that the first order terms in $\frac{1}{c}$ cancel and we are left
with the zeroth and second order terms which only partially cancel, as detailed in \ern{Fr2}.
The cancellation of first order terms is indeed remarkable as was pointed out by Feynman
 \cite{Feynman} with respect to the electromagnetic case.

$\vec F_N$ first introduced by Newton is attractive, however, the retardation force $\vec F_r$ can be either  attractive or repulsive. Newtonian force decreases as $\frac{1}{R^2}$,  however, ~the retardation force doe not depend on distance as long as the Taylor approximation given in Equation~(\ref{rhotay}) holds. Below~a certain distance, the~Newtonian force dominates, but~for larger distances the retardation force has the upper hand. Newtonian force can be neglected for distances significantly larger compared to the retardation distance:
\beq
 R \gg R_r \equiv c \Delta t
\label{Rr}
\enq
$\Delta t$ is a duration associated with the second order derivative of the density $\rho$. For~$R\ll R_r$, retardation can be neglected and only Newtonian forces need to be considered; this is the situation in the solar system.
As the galaxy accretes intergalactic gas, the galactic mass becomes larger thus $\dot{M}>0$;
however, the intergalactic gas is depleted, and thus the~rate at which the mass is accreted decreases resulting in $\ddot{M}<0$, hence we have an attractive retardation force.

One may claim that since for the galaxy $\ddot{M}<0$ and  the total mass is conserved it must be that $\ddot{M}>0$ for the matter outside the galaxy and thus retardation forces $\vec F_{ar}$ inside and outside the galaxy should cancel out. This derivation, however, is false because \ern{Fr2} is only valid when $\frac{R}{c}$ is small, it is certainly not small if the rest
of the universe outside the galaxy is taken into account. We shall later show by a detailed model that a retardation force exist regardless if one assumes the expansion of \ern{rhotay} or not.

As a final comment to this section please notice that
if we introduce dimensionless quantities using suitable dimensional constants, such that:
\beq
\tilde{\rho} \equiv \frac{\rho}{\rho_c}, \quad  \tilde{x} \equiv \frac{\vec x}{R_s},
\quad  \bar{t} = \frac{t}{\Delta t}
\label{dimles}
\enq
and define $\Lambda$ such that:
\beq
\Lambda = \int \tilde{\rho} d^3\tilde{x}
\label{dimles2}
\enq
we obtain that:
\beq
M = \int \rho d^3 x = \rho_c R_s^3 \int \tilde{\rho} d^3\tilde{x} = \Lambda \rho_c R_s^3.
\label{M}
\enq
Thus $\vec F_{ar}$ given in \ern{Fr2} can be written in terms of the Schwarzschild radius $r_s = \frac{2 G M}{c^2}$ as follows:
\beq
 \vec F_{ar} =  \frac{r_s}{ \Delta t^2}  \left[\frac{1}{4 \Lambda} \int  \frac{\partial^2 \tilde \rho}{\partial \bar{t}^2} \hat R d^3 \tilde x'\right].
\label{Far2}
\enq
the vector $\frac{1}{4 \Lambda} \int  \frac{\partial^2 \tilde \rho}{\partial \bar{t}^2} \hat R d^3 \tilde x'$ is dimensionless.

\section{Higher order terms}
\label{Hot}

Comparing equation (31) to equation (82) of \cite{YahalomSym} it follows that:
\beq
g^{(2)} (t) =  g^{(2)} (0) e^{\frac{t}{\tau}}
\label{g2t}
\enq
Hence:
\beq
g^{(n)} (t) =  g^{(2)} (0) \tau^{2-n} e^{\frac{t}{\tau}} =  g^{(2)} (t) \tau^{2-n}, \qquad n>2
\label{gnt}
\enq
And also:
\beq
\rho^{(n)} (\vec x, t) =  \rho^{(2)} (\vec x, t) \tau^{2-n}, \qquad M^{(n)} (t) =  M^{(2)} (t) \tau^{2-n}, \qquad n>2
\label{thont}
\enq
Thus according to \ern{phir} we have the following correction to the retardation potential:
\beq
\phi_{(n>2)}= -G \sum_{n=3}^{\infty} \frac{(-1)^n}{n! c^n \tau^{n-2}} \int R^{n-1} \rho^{(2)} (\vec x', t) d^3 x'.
\label{phir5}
\enq
The deviation from the second order approximation is more pronounced for large $r= |\vec x|$ for which $R = |\vec x - \vec x'| \simeq r$ which is the case we consider here, thus:
\beq
\phi_{(n>2)}\simeq  -G \sum_{n=3}^{\infty} \frac{(-1)^n}{n! c^n \tau^{n-2}} r^{n-1} \int  \rho^{(2)} (\vec x', t) d^3 x'
 = - \frac{G \ddot M(t) \tau^2}{r} \sum_{n=3}^{\infty} \frac{1}{n!} \left(\frac{-r}{c \tau} \right)^n.
\label{phir6}
\enq
Now using the well known identity:
\beq
\sum_{n=3}^{\infty} \frac{\alpha^n}{n!} = e^\alpha - (1+ \alpha + \frac{1}{2} \alpha^2)
\label{ident}
\enq
We may write \ern{phir6} as a closed expression instead of an infinite sum:
\beq
\phi_{(n>2)}\simeq - \frac{G \ddot M(t) \tau^2}{r} \left(e^{-\frac{r}{c \tau}} - 1 + \frac{r}{c \tau} - \frac{1}{2} \left(\frac{r}{c \tau}\right)^2\right).
\label{phir7}
\enq
For $r \ll c \tau$ it is quite clear that the term in the parenthesis of \ern{phir7} vanishes, since:
\beq
\lim_{\alpha \rightarrow 0} \sum_{n=3}^{\infty} \frac{\alpha^n}{n!} =\lim_{\alpha \rightarrow 0}\left(  e^\alpha - (1+ \alpha + \frac{1}{2} \alpha^2) \right) = 0.
\label{ident2}
\enq
Hence $\phi_{(n>2)}$ can be neglected if indeed $r \ll c \tau$ for the relevant measurements of the M33 rotation curve, that is up to about $r < 20$ kpc. Now $\tau$ is dependent according to equation (81) of \cite{YahalomSym}  on the density gradient of the intergalactic medium (IGM) and the typical velocity in this medium. Although those values are not known precisely we may assume that $v_z \sim 100$ km /s and the typical gradient is the same as the gradient of the optical disk luminosity that is $\frac{1}{k} \sim 0.1$ kpc. Thus $\tau \sim 10^6$ years, and $\tau c \sim 300$ kpc, making the second order approximation used so far reasonable.

{\bf Thus, we have clarified the domain and range of the current model i.e. what approximations are made and their validity. We underline that the developments of this paper are refinements of an existing model. In another paper it is shown that the retardation approach is valid despite the
relatively low velocity of matter: please see section 9 of \cite{YaRe3}.}

\section{Retardation beyond the Taylor Expansion}
\label{RbT}

Although, we have shown in the previous section that second order expansion is sufficient in the framework of a particular galactic model, it is worthwhile to explore the domain of validity of the retardation phenomena in general. We have shown in section \ref{REBN} that the retardation phenomena
is not important at distances which are short with respect to the retardation length $r \ll R_r$, hence we need only to consider retardation at distances of about $r \geq \frac{R_r}{10}$. But is there an upper distance limit? Indeed the expansion treatment is valid only up to distances of $r < R_{max}$, but this is a property of the Taylor expansion not of the retardation phenomena. We shall
now show that indeed there is an upper distance $R_{rul}$ beyond which the retardation phenomena is not important any more.

Let us look at the potential of \ern{phi} in the limit of large $r$ in which $r$ is
much bigger than a typical scale of the system: $r \gg R_s$. In this limit $R \simeq r$ and thus the potential of \ern{phi} can be written as:
\beq
\phi \simeq -G \int \frac{ \rho (\vec x', t-\frac{r}{c})}{r} d^3  x'
= -\frac{G}{r} \int  \rho (\vec x', t-\frac{r}{c}) d^3  x' =  -\frac{G M}{r},
\label{philargdis}
\enq
in the above $M$ defined in \ern{M} is the total mass of the system which is constant in time.
The same analysis can be repeated for a Newtonian potential:
\beq
\phi_N \simeq -G \int \frac{ \rho (\vec x', t)}{r} d^3  x'
= -\frac{G}{r} \int  \rho (\vec x', t) d^3  x' =  -\frac{G M}{r},
\label{philargdisN}
\enq
it follows that for large distances such that $r \gg R_s$  it makes not difference if we use a Newtonian or retarded potential, hence "dark matter" effects disappear. The above is only
true for an isolated system that does exchange mass with its environment. For a galaxy this will
include its "sphere of influence" which is not limited to the observable galaxy but includes also a surrounding from which the observable galaxy accretes mass from. Mathematically we may define:
\beq
\Delta \phi \equiv \phi  -\phi_N.
\label{delphi}
\enq
Thus:
\beq
\lim_{r->\infty} \Delta \phi  = \lim_{r->\infty} (\phi  -\phi_N) = 0,
\label{delphiinf}
\enq
indicating that there is an upper scale $R_{rul} \simeq 10 R_s$ above which $r>R_{rul}$ retardation
effects are unimportant. We can summarize the range of validity of the retardation phenomena by
the inequality:
 \beq
\frac{R_r}{10} < r < 10 R_s
\label{reta}
\enq
leading us to expect no retardation phenomena for systems in which:
\beq
R_r > 100 R_s.
\label{reta2}
\enq
To study the phenomena of retardation within a specific model we write the potential of \ern{phi} using dimensionless coordinates:
\beq
\phi = -G \frac{\rho_c R_s^3}{r} r \int \frac{\tilde \rho (\vec x', t-\frac{R}{c})}{R} d^3 \tilde x'
 =  - \frac{G M}{r}
 \frac{1}{\Lambda}\int \frac{r}{R} \tilde \rho (\vec x', t-\frac{R}{c})d^3 \tilde x'
\label{phidl}
\enq
in which we used \ern{M}. This leads to the definition:
\beq
\psi_r \equiv  \frac{1}{\Lambda}\int \frac{r}{R} \tilde \rho (\vec x', t-\frac{R}{c})d^3 \tilde x'
\qquad \Rightarrow \qquad  \phi = - \frac{G M}{r} \psi_r
\label{psir}
\enq
Similarly we define a Newtonian $\psi_N$:
\beq
\psi_N \equiv  \frac{1}{\Lambda}\int \frac{r}{R} \tilde \rho (\vec x', t)d^3 \tilde x'
\qquad \Rightarrow \qquad  \phi_N = - \frac{G M}{r} \psi_N.
\label{psiN}
\enq
Now according to \ern{philargdis} and \ern{philargdisN}:
\beq
\lim_{r->\infty} \psi_N =  \lim_{r->\infty} \psi_r  = 1.
\label{psilim}
\enq
I thus follows that:
\beq
\Delta \psi \equiv \psi_r  - \psi_N  \qquad \Rightarrow \qquad \lim_{r->\infty} \Delta \psi = 0
\label{delpsilim}
\enq
And also according to \ern{delphi}:
\beq
\Delta \phi = - \frac{G M}{r} \Delta \psi
\label{delphipsilim}
\enq

\section{A Specific Model}
\label{SM}

\begin{figure}[H]
\vspace{1cm}
\centering
\includegraphics[width= 0.7\columnwidth]{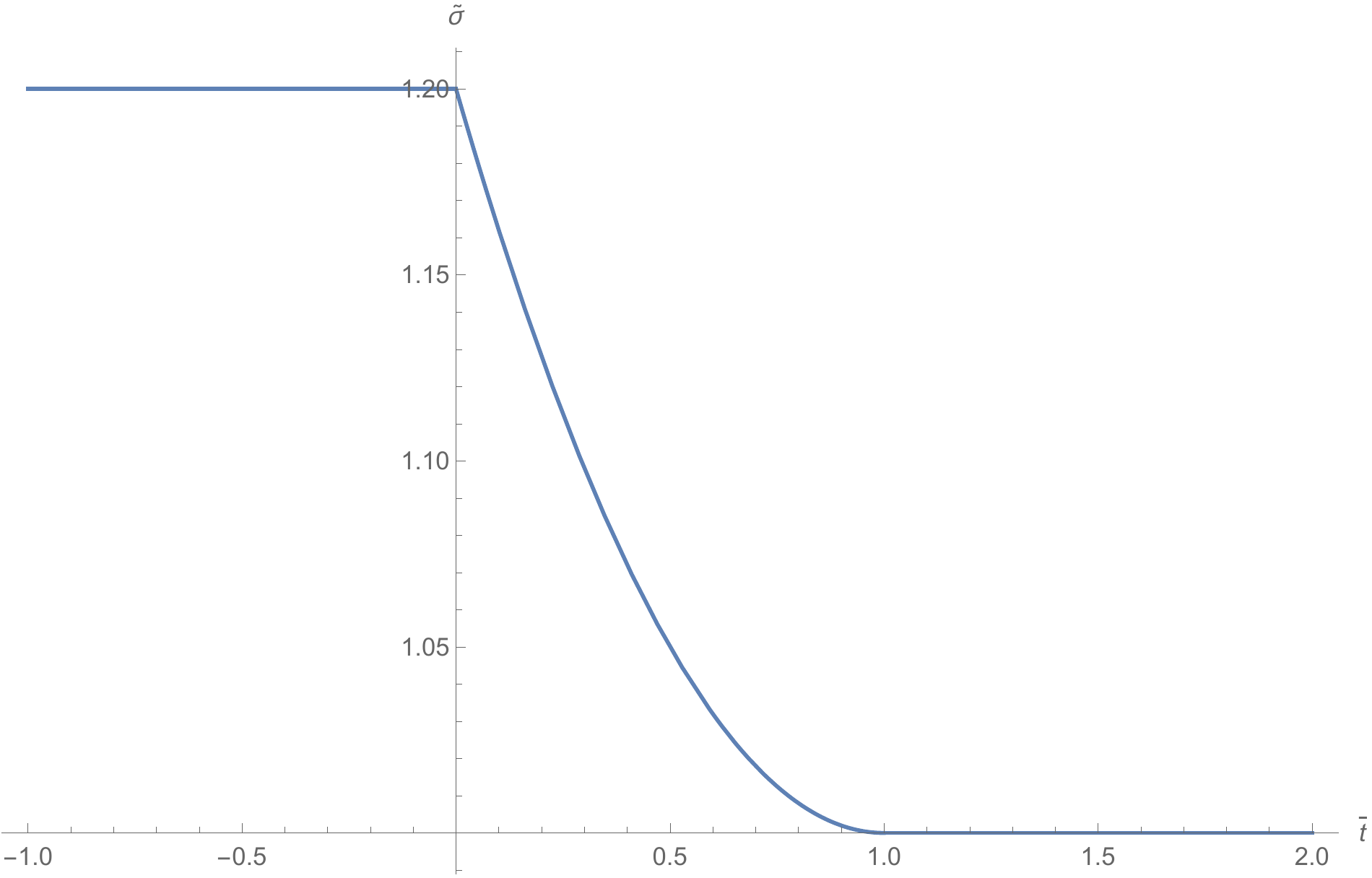}
 \caption{The profile width evolution.}
 \label{widthevo}
\end{figure}

\begin{figure}[H]
\vspace{1cm}
\centering
\includegraphics[width= 0.7\columnwidth]{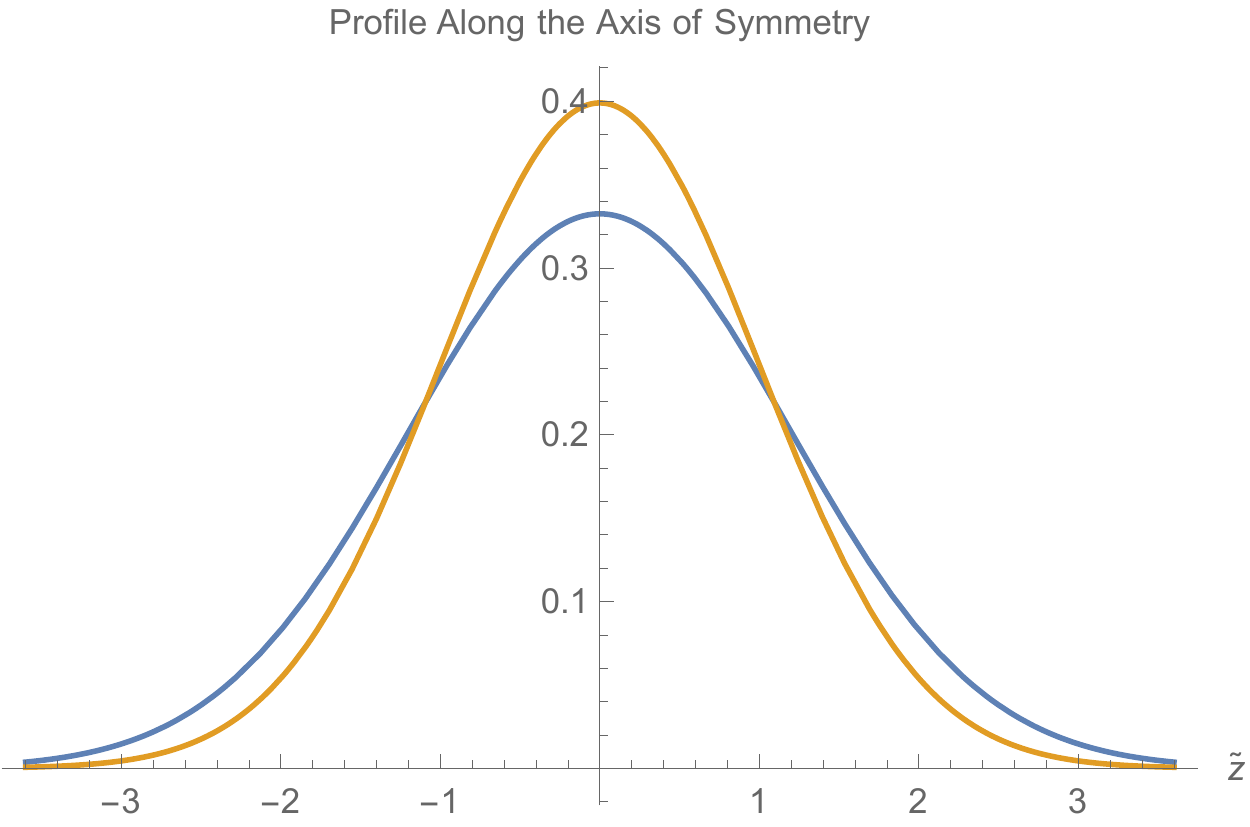}
 \caption{Axial density profile for $\bar{t}=0$ and $\bar{t}=1$.}
 \label{hprofg}
\end{figure}

\begin{figure}[H]
\includegraphics[width= 0.7\columnwidth]{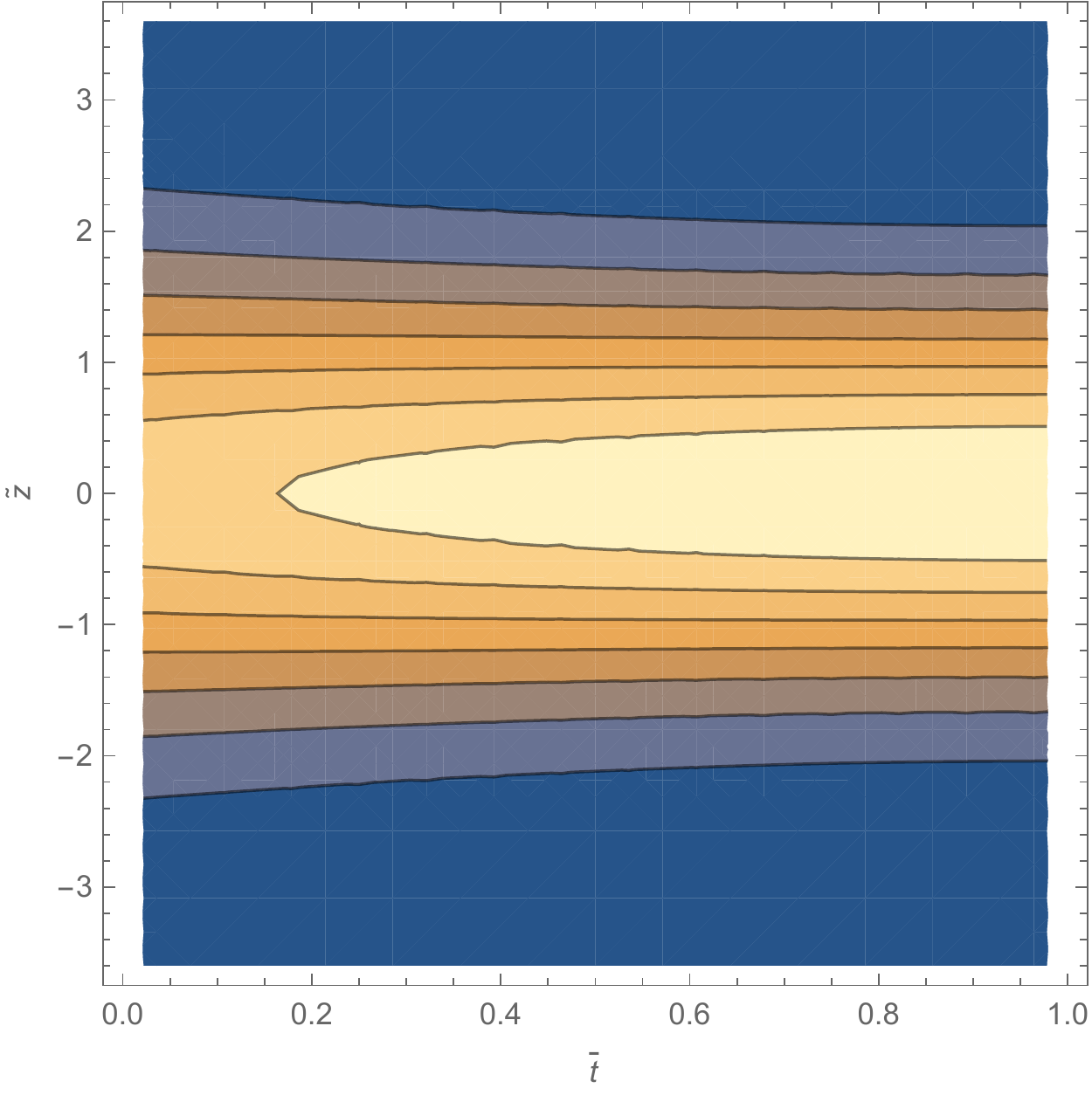}
\hspace{1cm}
\includegraphics[width= 0.1\columnwidth]{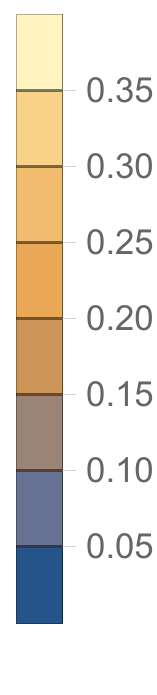}
 \caption{Axial density profile between $\bar{t}=0$ and $\bar{t}=1$.}
 \label{hprof2D}
\end{figure}

Let us now study the effects of retardation in the framework of a specific models:
\beq
\rho = \rho_c \tilde \rho, \qquad \tilde \rho = \Sigma (\vec x_\bot) h(z,t)
\label{denm}
\enq
in which $\Sigma (\vec x_\bot)$ is dependent on the coordinates in the transversal direction $\vec x_\bot$ and $h$ is dependent on the time $t$ and on the coordinates along the axial direction $z$.
We will further assume that:
\beq
\Sigma (\vec x_\bot) = \delta (\tilde x) \delta (\tilde y)
\label{transprof}
\enq
in which $\delta$ is a Dirac delta function. It is clear that any profile can be constructed from a weighted sum of $\Sigma (\tilde x - \tilde x_i, \tilde y- \tilde y_i)$ located at different $(\tilde x_i,\tilde y_i)$. For the axial direction we assume a Gaussian profile:
 \beq
 h(z,t) = \frac{R_s}{\sqrt{2 \pi} \sigma(t)} e^{-\frac{z^2}{2 \sigma(t)^2}}.
\label{hprof}
\enq
The time dependence is given through the width of the Gaussian profile which assumes the following form:
\beq
\sigma(t) = \left\{ \begin{array}{cc}
                     \sigma_i & \bar{t} \leq 0  \\
                     \sigma_i + (\sigma_f - \sigma_i) \bar{t} (2- \bar{t}) & 0 < \bar{t}<1 \\
                     \sigma_f & \bar{t} \geq 1
                   \end{array}
  \right. \qquad \bar{t} \equiv \frac{t}{t_f}
\label{sigt}
\enq
$t_f$ is a typical time scale, $\sigma_i$ is the initial time width before a change takes place
and $\sigma_f$ is the resulting time width after the change took place. We shall define
\beq
\tilde \sigma(t) = \frac{\sigma(t)}{R_s}, \qquad
\tilde \sigma_i = \frac{\sigma_i}{R_s}, \qquad
\tilde \sigma_f = \frac{\sigma_f}{R_s}.
\label{sigt2}
\enq
Hence:
 \beq
 h(z,t) = \frac{1}{\sqrt{2 \pi} \tilde \sigma(t)} e^{-\frac{\tilde z^2}{2 \tilde \sigma(t)^2}}.
\label{hprof2}
\enq
Choosing conveniently $\sigma_f = R_s$ and $\sigma_i = 1.2 R_s$ we obtain:
\beq
\tilde \sigma(t) = \left\{ \begin{array}{cc}
                    1.2 & \bar{t} \leq 0  \\
                     1.2  - 0.2 \bar{t} (2- \bar{t}) & 0 < \bar{t}<1 \\
                     1.0 & \bar{t} \geq 1
                   \end{array}
  \right.
\label{sigt3}
\enq
The profile width evolution
is depicted in figure \ref{widthevo}.
The axial density profile $h(z,t)$ is depicted for $\bar{t}=0$ and $\bar{t}=1$
in figure \ref{hprofg}, and for any time in between in figure \ref{hprof2D}
We note that for the current profile the mass $M$ is constant although the
density profile is time dependent, we also note that for the current density profile
we have $ \Lambda = 1$ according to \ern{dimles2}.

\begin{figure}[H]
\centering
\includegraphics[width= 0.7\columnwidth]{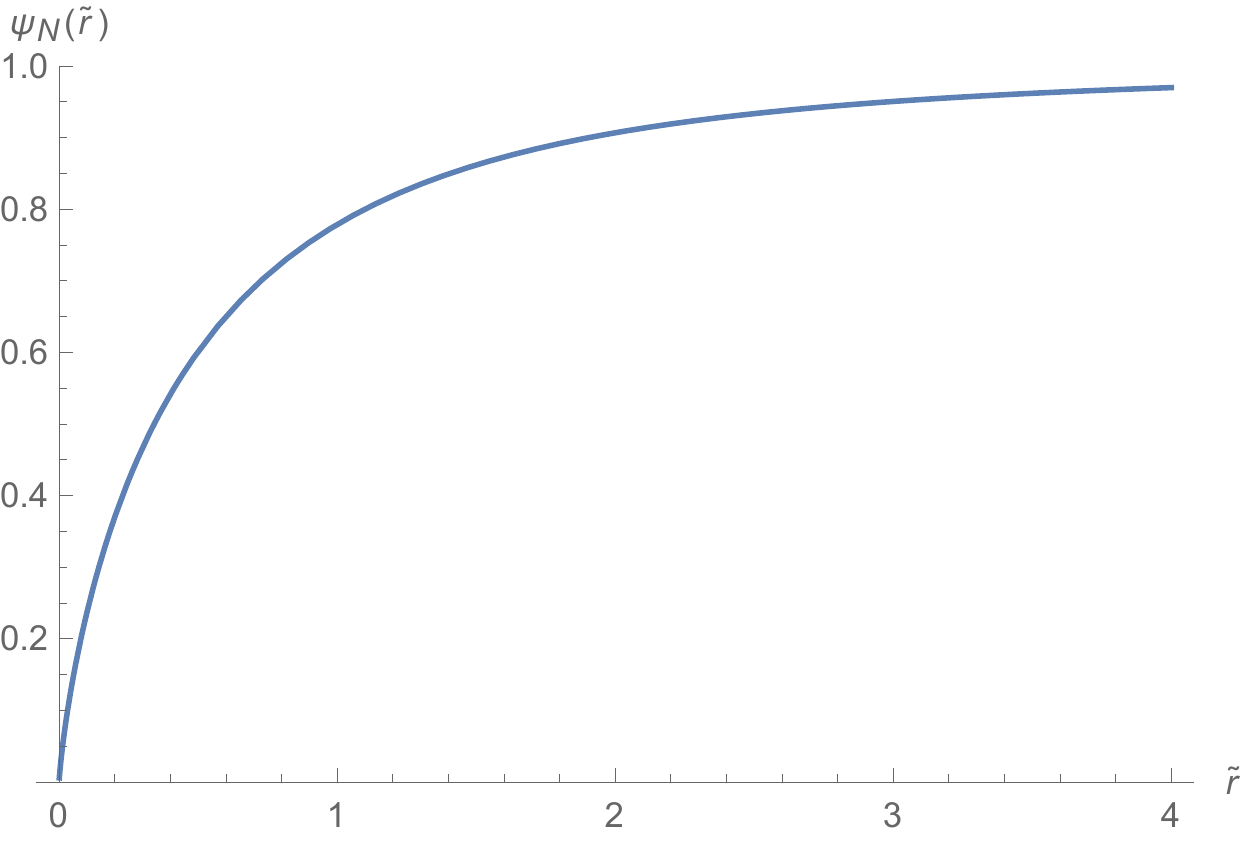}
 \caption{The function $\psi_N (\tilde r,0.5)$, it is clear that $\psi_N$ approaches $1$ asymptotically.}
 \label{psiNpl}
\end{figure}

\begin{figure}[H]
\centering
\includegraphics[width= 0.7\columnwidth]{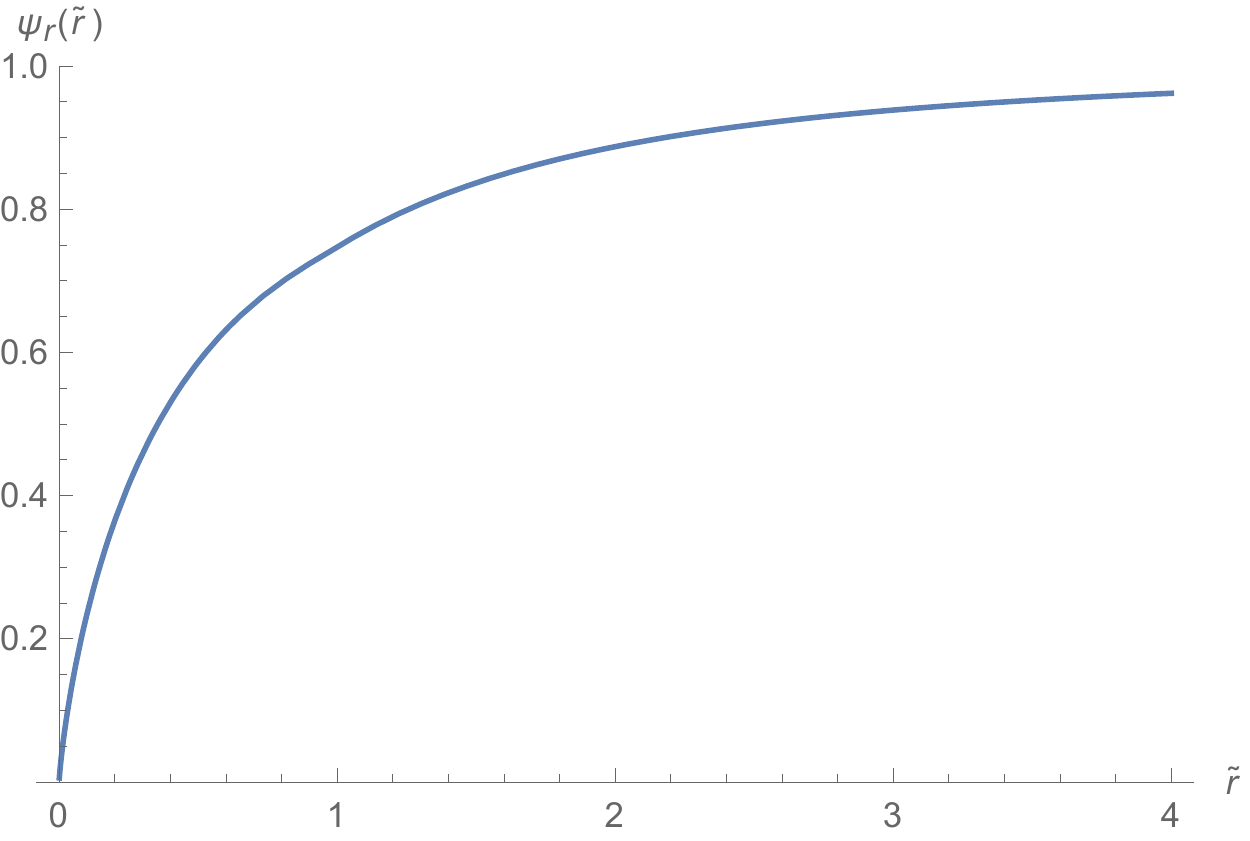}
 \caption{The function $\psi_r (\tilde r,0.5)$, it is clear that $\psi_r$ approaches $1$ asymptotically.}
 \label{psirpl}
\end{figure}

\begin{figure}[H]
\centering
\includegraphics[width= 0.7\columnwidth]{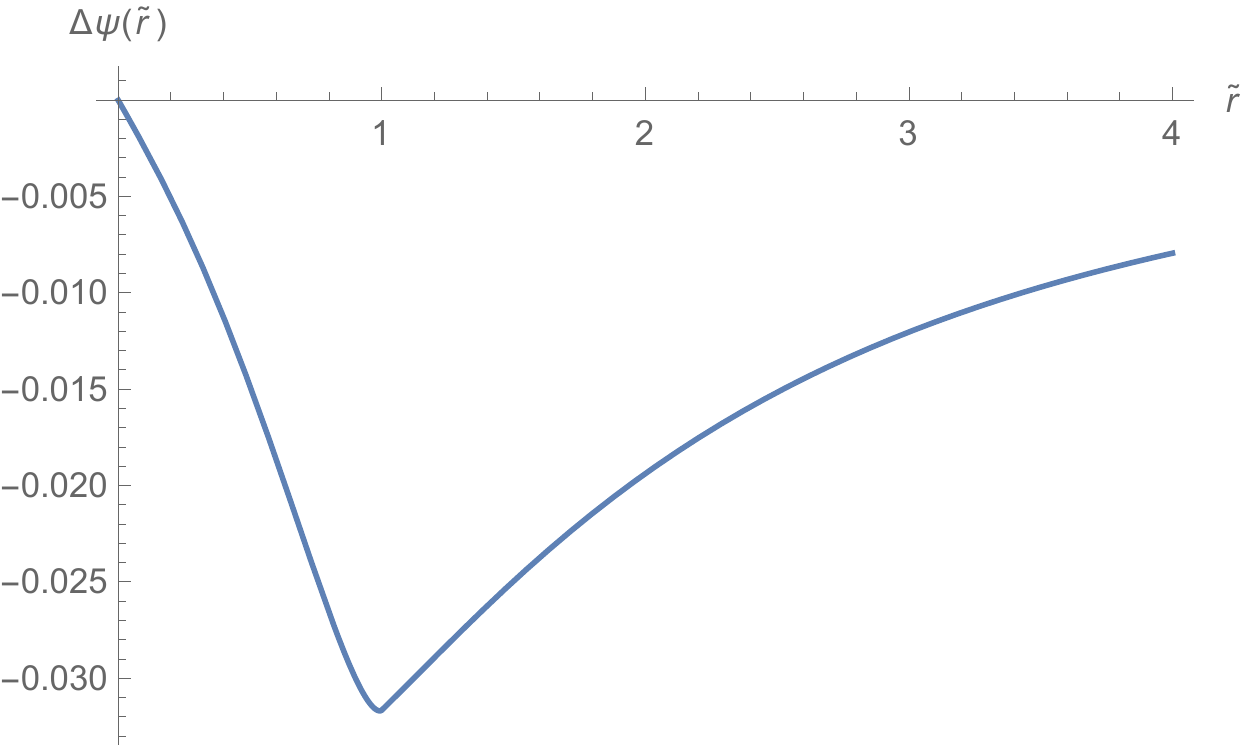}
 \caption{The function $\Delta \psi (\tilde r,0.5)$ depicted up to $ \tilde r = 4$.}
 \label{psidif1}
\end{figure}

\begin{figure}[H]
\centering
\includegraphics[width= 0.7\columnwidth]{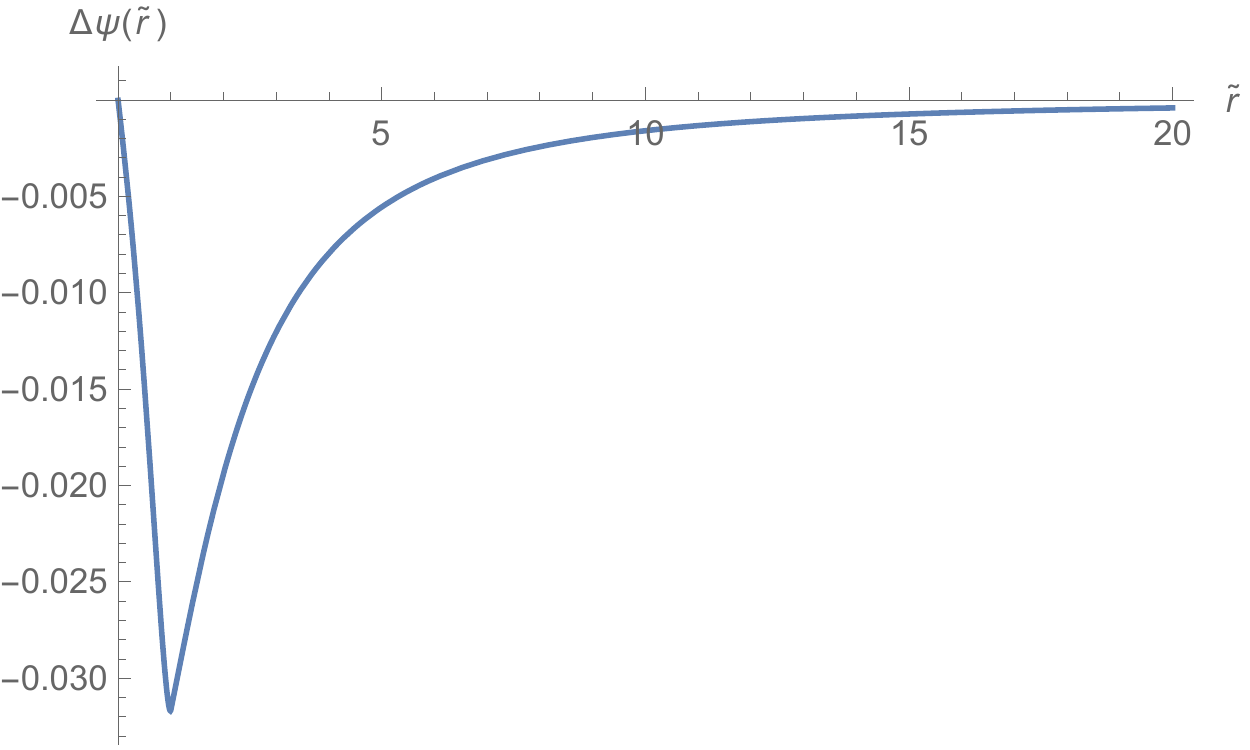}
 \caption{The function $\Delta \psi (\tilde r,0.5)$ depicted up to $ \tilde r = 20$.}
 \label{psidif2}
\end{figure}

We are now in a position to calculate $\psi_N$ and $\psi_r$ using \ern{denm}, \ern{transprof}
and \ern{hprof2}. We shall choose a point of distance $r$ from the origin along the $x$ axis located at the $z=0$ plane to evaluate $\psi$, since the density profile is cylindrically symmetric every point in the $z=0$ plane of distance $r$ to the origin is equivalent to any other. Taking into account \ern{psiN} we have:
\beq
\psi_N (\tilde r,\bar t) = \int \frac{\tilde r}{ \tilde R} h (\tilde z', \bar t)d \tilde z',
\qquad \tilde R = \sqrt{\tilde z'^2 + \tilde r^2}.
\label{psiNeva}
\enq
The function $\psi_N (\tilde r,0.5)$ is depicted in figure \ref{psiNpl} in which it is clear that $\psi_N$
asymptotically approaches $1$.
Taking into account \ern{psir} we have:
\beq
\psi_r (\tilde r,\bar t) = \int \frac{\tilde r}{ \tilde R} h (\tilde z', \bar t-
\frac{R}{c t_f} )d \tilde z'.
\label{psireva1}
\enq
Hence we have another length scale $R_{s2} = c t_f$ which is conveniently chosen to be $R_{s2} = 2  R_{s}$. The function $\psi_r (\tilde r,0.5)$ is depicted in figure \ref{psirpl} in which it is clear that $\psi_r$  also asymptotically approaches $1$.
The difference between the two functions $\Delta \psi$ is defined in \ern{delpsilim} and depicted
in figure \ref{psidif1}. It is clear that the difference exist and does not depend on a Taylor expansion, it also clear that this difference may be approximated by a simple function to about
$ \tilde r = 1$, or starting at $ \tilde r = 1$ and using a simple function to describe the function outwards which would work for at least $ \tilde r = 4$. To see that the function $\Delta \psi$ indeed approached $0$ asymptotically as predicted by \ern{delpsilim} the $\tilde r$ axis is extended to $ \tilde r = 20$ as depicted in figure \ref{psidif2}.

\begin{figure}[H]
\centering
\includegraphics[width= 0.7\columnwidth]{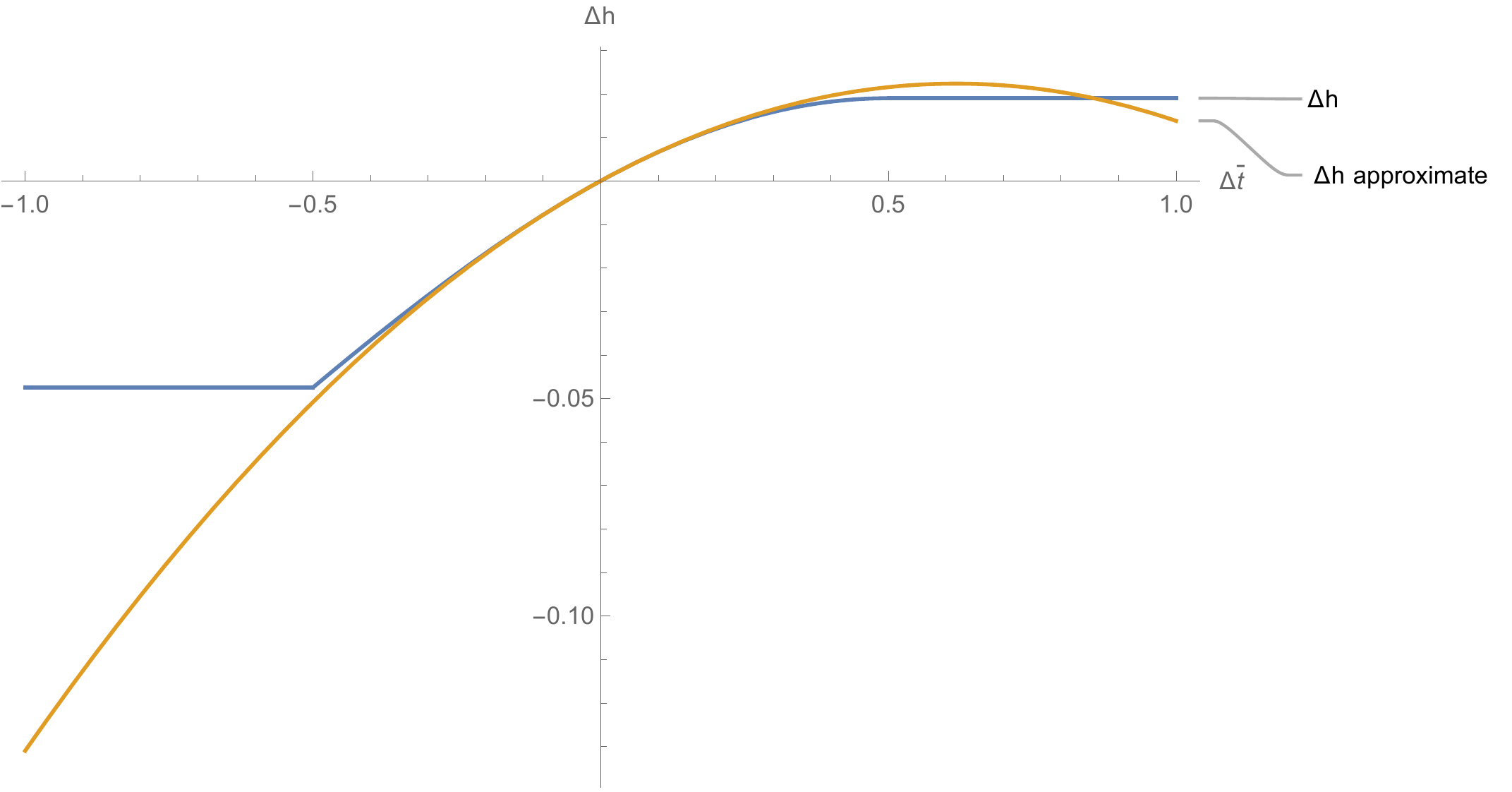}
 \caption{The functions $\Delta h(0, 0.5, \Delta \bar{t})$ and $\Delta h_a(0, 0.5,\Delta \bar{t})$.}
 \label{hafig}
\end{figure}

\begin{figure}[H]
\centering
\includegraphics[width= 0.7\columnwidth]{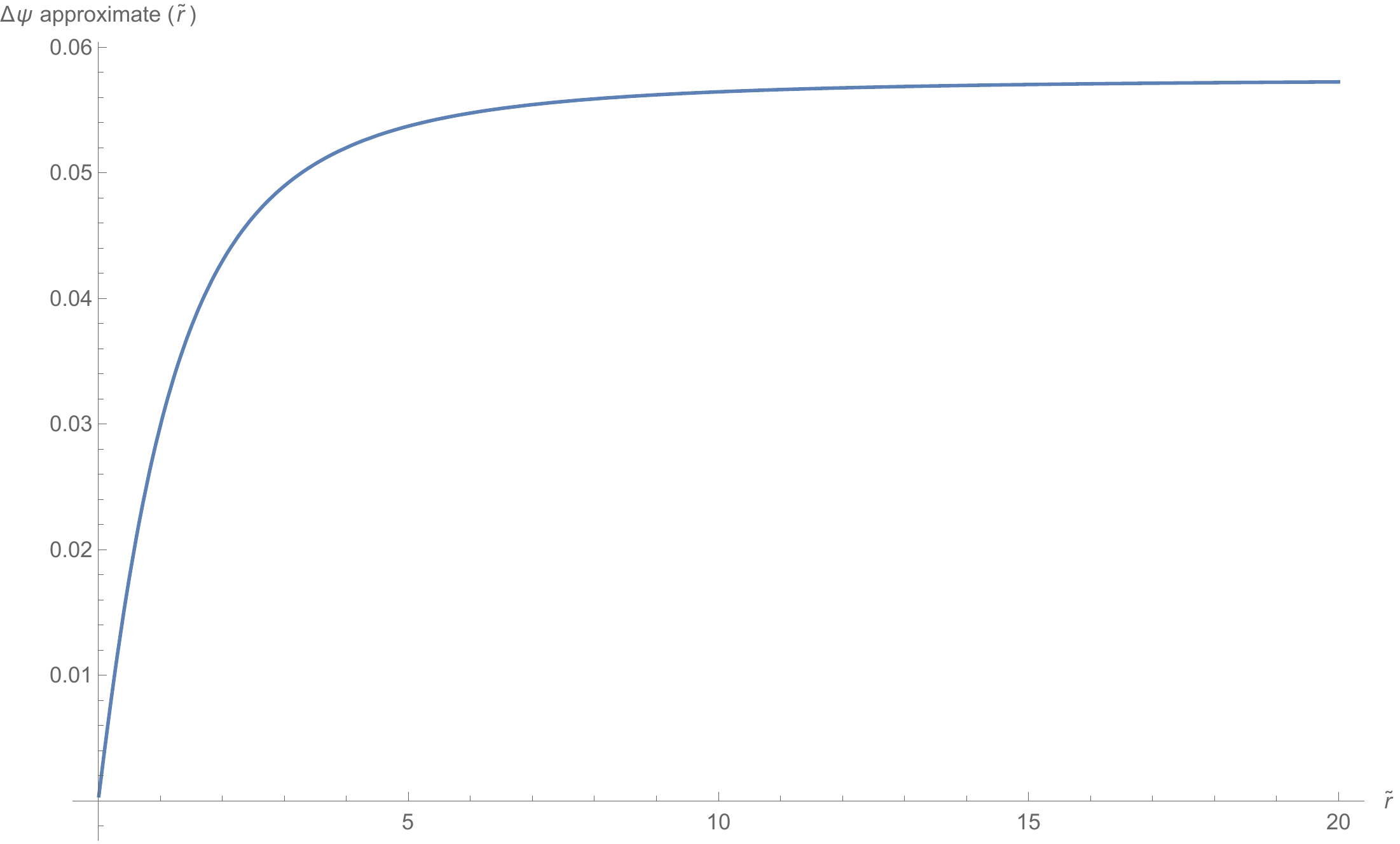}
 \caption{The asymptotic behaviour of $\Delta \psi_{aw}(\tilde r, \bar t)$ is clearly not appropriate.}
 \label{delpsiaw}
\end{figure}

To conclude this section we shall discuss the applicability of a Taylor expansion of the type used
in \ern{rhotay}. In the current context, we study the function $h(\tilde z, \bar t)$, that is
we would like to know how good is the approximation:
\beq
\Delta h(\tilde z, \bar t, \Delta \bar{t}) \equiv
h(\tilde z, \bar t+ \Delta \bar{t}) - h(\tilde z, \bar t) \simeq \Delta h_a(\tilde z, \bar t, \Delta \bar{t}),
\qquad
\Delta h_a(\tilde z, \bar t, \Delta \bar{t}) \equiv  \Delta \bar{t} h^{(1)}(\tilde z, \bar t)
+ \frac{1}{2}  \Delta \bar{t}^2 h^{(2)}(\tilde z, \bar t)
\label{haprox}
\enq
The numerical evaluation of the functions $\Delta h(\tilde z, \bar t,\Delta \bar{t})$ and $\Delta h_a(\tilde z, \bar t,\Delta \bar{t})$ are depicted in figure \ref{hafig} for the plane $z=0$ and for the time $\bar t = 0.5$. It is easy to see that the approximation is indeed a good one for $
|\Delta \bar t| <0.5$, otherwise the approximation is not valid especially for a negative
$\Delta \bar t$ which is our main concern in retardation theory.
The next step is to set the retardation delay $\Delta \bar{t} = - \frac{R}{c t_f} =- \frac{R}{R_{s2}} = - \frac{\tilde R}{\tilde R_{s2}}$ and calculate the approximate form of $\Delta \psi$:
\beq
\Delta \psi(r, \bar t) = \psi_r(r, \bar t)  - \psi_N(r, \bar t)
=  \int \frac{\tilde r}{ \tilde R} \Delta h(\tilde z', \bar t, - \frac{\tilde R}{\tilde R_{s2}})d \tilde z' \simeq \Delta \psi_a(r, \bar t).
\label{delpsidelh}
\enq
We may try the following approximation:
\beq
\Delta \psi_{aw}(r, \bar t) \equiv
 \int_{-\infty}^{\infty} \frac{\tilde r}{\tilde R} \Delta h_a(\tilde z', \bar t, - \frac{\tilde R}{\tilde R_{s2}})d \tilde z' = -\frac{\tilde r}{\tilde R_{s2}} \int_{-\infty}^{\infty} h^{(1)}(\tilde z', \bar t) d \tilde z'
 +\frac{\tilde r}{2 \tilde R_{s2}}\int_{-\infty}^{\infty} \frac{\tilde R}{\tilde R_{s2}} h^{(2)}(\tilde z', \bar t) d \tilde z'.
\label{delpsidelh2}
\enq
However, notice that for many points in the integration domain $|\frac{\tilde R}{\tilde R_{s2}}|$
is not smaller than $\frac{1}{2}$ hence $\Delta h_a$ is not a valid approximation. Moreover,
notice that:
\beq
\int_{-\infty}^{\infty} h^{(1)}(\tilde z', \bar t) d \tilde z'
 = \frac{\partial}{\partial \bar{t}} \int_{-\infty}^{\infty} h(\tilde z', \bar t) d \tilde z'
 = \frac{\partial}{\partial \bar{t}} 1 = 0.
\label{firsttv}
\enq
Hence:
\beq
\Delta \psi_{aw}(\tilde r, \bar t) =
\frac{\tilde r}{2 \tilde R_{s2}}\int_{-\infty}^{\infty} \frac{\tilde R}{\tilde R_{s2}} h^{(2)}(\tilde z', \bar t) d \tilde z'.
\label{delpsidelh3}
\enq
However, $\Delta \psi_{aw}(r, \bar t)$ does not have the asymptotic property $\lim_{r->\infty} \Delta \psi_{aw}(r, \bar t) = 0 $ described in \ern{delpsilim} as can be seen by a numerical
evaluation presented in figure \ref{delpsiaw}. We conclude that $\Delta \psi_{aw}$ is {\bf not}
an adequate approximation of $\Delta \psi$.
Notice, however, that this problem can be somewhat elevated if one integrates close enough
to the galactic plane $z=0$ such that $|\frac{\tilde R}{\tilde R_{s2}}|<0.5$ and consider
small $\tilde r$. Indeed, most of the mass in centered close to the galactic plane. After some numerical experimentation we obtained a reasonable approximation integrating in the range: $[-0.25 \tilde \sigma_i,0.25 \tilde \sigma_i]$:
\ber
\Delta \psi_{ac}(r, \bar t) &\equiv&
 \int_{-0.25 \tilde \sigma_i}^{0.25 \tilde \sigma_i} \frac{\tilde r}{\tilde R} \Delta h_a(\tilde z', \bar t, - \frac{\tilde R}{\tilde R_{s2}})d \tilde z'
 \nonumber \\
 &=& -\frac{\tilde r}{\tilde R_{s2}} \int_{-0.25 \tilde \sigma_i}^{0.25 \tilde \sigma_i} h^{(1)}(\tilde z', \bar t) d \tilde z'
 +\frac{\tilde r}{2 \tilde R_{s2}}\int_{-0.25 \tilde \sigma_i}^{0.25 \tilde \sigma_i} \frac{\tilde R}{\tilde R_{s2}} h^{(2)}(\tilde z', \bar t) d \tilde z'.
\label{delpsidelh2c}
\enr
\begin{figure}[H]
\centering
\includegraphics[width= 0.7\columnwidth]{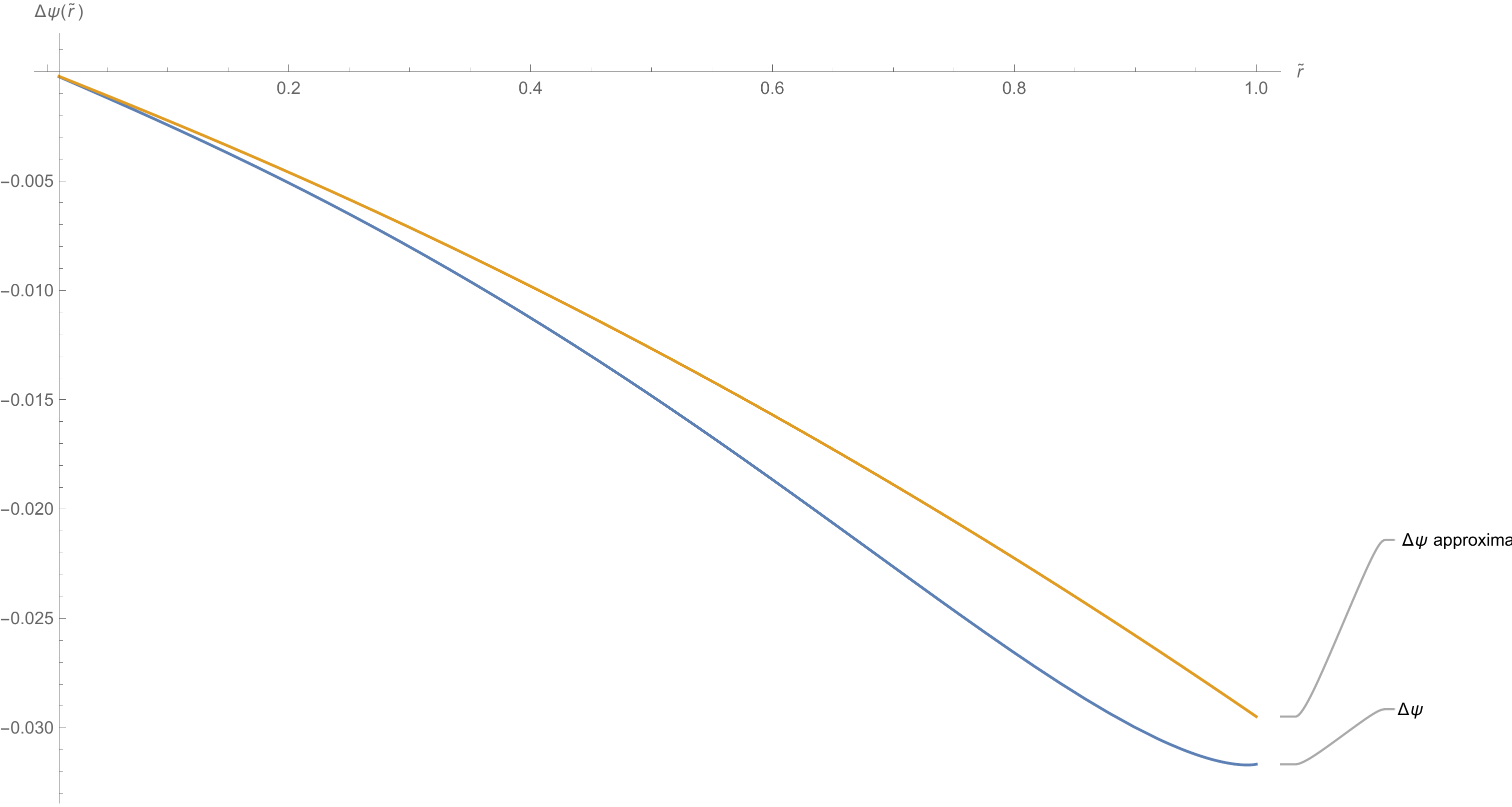}
 \caption{$\Delta \psi(\tilde r,0.5)$ and $\Delta \psi_{ac}(\tilde r,0.5)$ for $\tilde{r} <1$.}
 \label{delpsiac}
\end{figure}
As a final comment we stress again that $R_s$ is the scale of the "galactic sphere of influence" that is the typical dimension of the region with which the galaxy exchanges mass and not necessarily the galactic radius. Thus the domain $\tilde r < 1$ may stretch far beyond the galaxy itself.

\section{Conclusions}

The phenomena of retardation is ubiquitous in physics, and follows directly from the Lorentz symmetry group. Hence, any system that is invariant under the Lorentz transformation will exhibit retardation phenomena. Those include physical systems related to classical electromagnetism ~\cite{Tuval,YahalomT,Yahalom3,Yahalom4} General Relativity \cite{YaRe1,ge,YaRe2,YahalomSym}, but also to other Lorentz invariant theories such as conformal gravity~\cite{Mannheim0,Mannheim1,Mannheim2}.

Dark matter being a major candidate to explain galactic rotation curves has only a slim chance to being found, given that accelerator experiments, as using the Large Hadron Collider was unable to find any super symmetric particles, not only of the community's favorite form of dark matter, but~also the form of it that is mandated in string theory, a~theory that also suggests a quantized version of Einstein~gravity.

We have shown that at least on the galactic scale dark matter is not needed \cite{YaRe1,ge,YaRe2,YahalomSym,Wagman}, as the dynamics can be explained by a retarded gravitational potential when a near field approximation is used. We remark that the analysis of
far field leading to gravitational waves \cite{Einstein2} was corroborated in recent years by observations \cite{Taylor,Castelvecchi}.

A justification for the second order Taylor series approximation which we used in previous works is given here for the first time, showing that indeed higher order terms can be safely neglected.

We also show that one may discuss retardation phenomena without relying on a Taylor expansion.

\authorcontributions{This paper has a single author, which has done all the reported work.}

\funding{This research received no funding.}

\acknowledgments{The author wishes to thank his former student, Dr. Michal Wagman, for being critical regarding retardation theory. Much of this paper is a result of her raising various questions on the validity of the second order approximation which was used in \cite{YahalomSym}.
The same doubts were raised by my colleague and friend Prof. Yosef Pinhasi, I would like to thank him as well for a critical discussion.}

\conflictsofinterest{The author declares no conflict of~interest.}

%%%%%%%%%%%%%%%%%%%%%%%%%%
\reftitle{References}

\publishersnote{MDPI stays neutral with regard to jurisdictional claims in published maps and institutional affiliations.}
\end{document}